\documentclass[showpacs,preprintnumbers,amsmath,amssymb,floatfix,eqsecnum]{revtex4}

\usepackage{color}   
\usepackage{graphicx}
\usepackage{dcolumn}
\usepackage{bm}

\begin{document}

\def\bbox#1{\hbox{\boldmath${#1}$}}
\def\blambda{{\hbox{\boldmath $\lambda$}}}
\def\eeta{{\hbox{\boldmath $\eta$}}}
\def\bxi{{\hbox{\boldmath $\xi$}}}
\def\bzeta{{\hbox{\boldmath $\zeta$}}}


\title{ Lectures on Landau Hydrodynamics\footnote{Lectures presented
    at the Helmholtz International Summer School, Bogoliubov
    Laboratory of Theoretical Physics, JINR, Dubna, July 14-26, 2008}
    }

\author{Cheuk-Yin Wong\footnote{wongc@ornl.gov}}

\affiliation{Physics Division, Oak Ridge National Laboratory, Oak
Ridge, TN  37831}

\date{\today}

\begin{abstract}

Landau hydrodynamics is a plausible description for the evolution of
the dense hot matter produced in high-energy heavy-ion collisions.  We
review the formulation of Landau hydrodynamics to pave the way for its
application in high-energy heavy-ion collisions.  It is found that
Landau's rapidity distribution needs to be modified to provide a
better quantitative description.  In particular, the rapidity
distribution in the center-of-mass system should be more appropriately
given as $dN/dy \propto \exp \{ \sqrt{y_b^2-y^2}\}$, where $y_b=\ln
\{\sqrt{s_{NN}}/m_p\}$ is the beam nucleon rapidity, instead of
Landau's original result of $dN/dy({\rm Landau}) \propto \exp \{
\sqrt{L^2-y^2}\}$ where $L=\ln \{\sqrt{s_{NN}}/2m_p\}$.  The modified
distribution is compared with the Landau distribution and experimental
data.  It is found that the modified distribution agrees better with
experimental $dN/dy$ data than the Landau distribution and it differs
only slightly from the Landau Gaussian distribution $dN/dy({\rm
Landau~Gaussian}) \propto \exp \{ -y^2/2L\}$.  Past successes of the
Gaussian distribution in explaining experimental rapidity data arises,
not because it is an approximation of the original Landau
distribution, but because it is in fact a close representation of the
modified distribution.

\end{abstract}


\maketitle


\large
 \section {\bf 
Introduction}
\vspace{0.3cm}

The evolution of the dense hot matter produced in high-energy
heavy-ion collisions is a problem of intrinsic interest.  In many
problems in high-energy heavy-ion collisions, such as in the
investigation of the fate of a heavy quarkonium in quark-gluon plasma
and the interaction of a jet with the produced medium, it is desirable
to have a good description of the dynamics of the matter produced in
the collision.  For these problems, Landau hydrodynamics
\cite{Lan53,Bel56} furnishes a plausible description for the evolution
of an assembly of dense matter at a high temperature and pressure. Its
dynamics during the first stage of one-dimensional longitudinal
expansion can be solved exactly 
\cite{Kha54,Bel56,Bel56a,Ama57,Sri92,Ham04,Pra07,Bia07,Cso07,Cso08,Beu08}.
The one-dimensional longitudinal expansion problem admits an
approximate solution that is applicable to the bulk part of the fluid.
The subsequent three-dimensional motion can be solved approximately to
give rise to predictions that come close to experimental data
\cite{Mur04,Ste05,Ste07,Bus04}.  A thorough understanding and critical
tests of the model will make it a useful tool for the description of
the evolution of the produced dense matter.  We therefore review the
formulation of Landau hydrodynamics to pave the way for its
quantitative applications in high-energy heavy-ion collisions.

Common to both Landau hydrodynamics and Bjorken hydrodynamics is the
basic assumption that in the dense hot matter produced in high-energy
heavy-ion collisions, the density of the quanta of the medium (either
quarks and gluons or the hadrons) is so high that the mean-free paths
of the constituents are short, and a state of local thermal
equilibrium can be maintained through out.  Such a condition may be at
hand in collisions at very high energies.  Under the circumstances,
the dynamics of the system will be determined by the equation of
hydrodynamics possessing local thermal equilibrium.  As we are dealing
with high temperatures and high densities for which the temperature is
of the order of the rest mass of the quanta, we need to examine the
problem with relativistic hydrodynamics.

The major difference between Landau hydrodynamics and Bjorken
hydrodynamics is in the nature of the flow.  In Bjorken hydrodynamics,
boost invariance is assumed so that the energy density $\epsilon$ is
independent of the rapidity.  It is a function of the proper time
$\tau$ only, $\epsilon_{\rm Bjorken}(\tau)$.  However, in Landau
hydrodynamics, no such an assumption is made, and the energy density
is a local function of spatial and temporal coordinates,
$\epsilon_{\rm Landau}(t,z,x,y)$.

We can consider collisions of two equal-size nuclei of diameter $a$.
The disk of initial configuration in the center-of-mass system has a
longitudinal thickness $\Delta$ given by
\begin{eqnarray}
\label{eq11}
\Delta = a /\gamma,
\end{eqnarray}
where $\gamma$ is the Lorentz contraction factor
\begin{eqnarray}
\gamma= \frac{\sqrt{s_{_{NN}}}}{2m_p},
\end{eqnarray}
 $\sqrt{s_{_{NN}}}/2$ is the center-of-mass energy per nucleon, and
$m_p$ is the proton mass.  The dimensions of the disk are not
symmetrical in the longitudinal and transverse directions.  The
initial configuration can be depicted as in Fig.\ 1 where it can be
represented as an elliptical disk with a thickness $\Delta$ and major
diameters $a_x$ and $a_y$, with the reaction plane lying on the
$x$-$z$ plane.  Depending on the impact parameter, the dimensions of
the disk obey $a_x \le a_y \le a$, and the azimuthal radius $a_\phi/2$
depends on the azimuthal angle $\phi$ measured from the $x$-axis.  For
a central collision, $a_x = a_y = a$.

\begin{figure} [h]
\includegraphics[angle=0,scale=0.50]{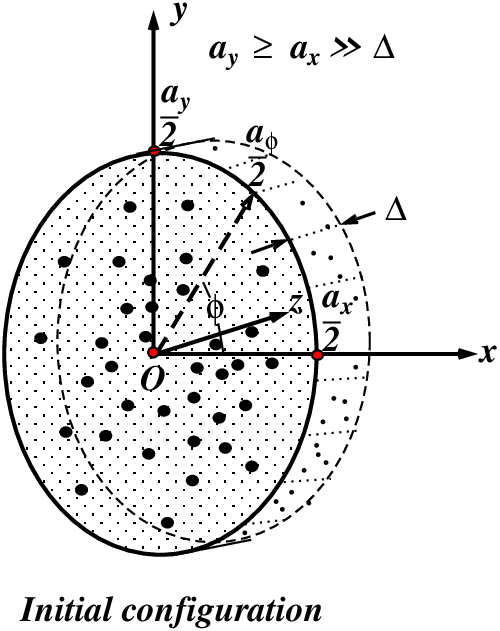}
\vspace*{0.0cm} 
\caption{Initial configuration in the collision of two heavy nuclei in
  the center-of-mass system.  The region of nuclear overlap consists
  of a thin disk of thickness $\Delta$ along the longitudinal
  $z$-axis. The reaction plane is designated to lie on the $x$-$z$
  plane, and the transverse radii are $a_x/2$ and $a_y/2$.}
\end{figure}

The matter in the disk can be described as a fluid with an energy
density field $\epsilon$, and a 4-velocity field $u^\mu$.  In the
center-of-mass system to which we focus our attention, the disk of
matter can be considered to be initially nearly at rest.  The density
of the produced matter and its associated pressure are very large,
whereas the spatial dimensions of the fluid are very small.  As a
consequence, large pressure gradients are generated.  The generated
pressure gradients lead to the subsequent expansion of the system in
all directions.  

In highly relativistic collisions to which we focus our attention, the
center-of-mass energy per nucleon is much greater than the nucleon
rest mass, $(\sqrt{s_{NN}}/2) \gg m_p$.  We have then $\Delta \ll a_x
\le a_y$ and $\nabla_z p \gg \nabla_x p \ge \nabla_y p$.  The pressure
gradient in the longitudinal $z$ direction far exceeds those along the
transverse $x$ and $y$ directions.  The expansion along the
longitudinal direction proceeds much faster than the transverse
expansions. Among the two transverse directions, the expansion along
the $x$ direction proceeds slightly faster than the expansion along
the $y$ direction, leading to the presence of the elliptic flow.

The large difference of the expansion speeds in the longitudinal and
transverse directions allows Landau to split the evolution dynamics
into two stages.  In the first stage, one considers a one-dimensional
longitudinal expansion in conjunction with a simultaneous but slower
transverse expansion.  The second stage occurs when the transverse
displacement exceeds the initial transverse dimensions so that the
transverse force acting on the fluid element become so small as to be
negligible.  Thus, in the second stage, the fluid element will freeze
its rapidity to stream out in a conic flight with a fixed polar angle.
The magnitude of the transverse displacement decreases with 
increasing rapidity magnitudes. Therefore different parts of the fluid
will enter from the first stage to the second stage at different
times.  Those fluid elements with the smallest rapidity magnitude will
enter the second stage the earliest while those elements with the
largest rapidity magnitude will enter the second stage the latest.  By
matching the solutions for the first and the second stages, the
dynamics of the system can be followed, up to the end point of the
hydrodynamical evolution.

Recent comparison with experimental data indicates that the Landau
hydrodynamical model yields results that agree with experiment.  In
Refs. \cite{Mur04,Ste05,Ste07,Car73}, quantitative analyses use an
approximate form of the Landau rapidity distribution
\cite{Lan53,Bel56},
\begin{eqnarray}
\label{gau}
dN/dy\propto \exp\{-y^2/2L\},
\end{eqnarray}
where $L$ is the logarithm of the Lorentz contraction factor $\gamma$ 
\begin{eqnarray}
L=\ln \gamma= \ln (\sqrt{s_{NN}}/2m_p).
\end{eqnarray}
This Gaussian form of the rapidity distribution gives theoretical
rapidity widths that agree with experimental widths for many different
particles in central AuAu collisions, to within 5 to 10\%, from AGS
energies to RHIC energies \cite{Mur04,Ste05,Ste07}.  The Landau
hydrodynamical model also gives the correct energy dependence of the
observed total multiplicity, and it exhibits the property of
``limiting fragmentation'' at forward rapidities \cite{Ste05,Ste07}.
A similar analysis in terms of the pseudorapidity variable $\eta$ at
zero pseudorapidity has been carried out in \cite{Sar06}.

The successes of these quantitative and qualitative analyses
indicate that Landau hydrodynamics can be a reasonable description.
However, they also raise many unanswered questions.  Firstly, the
original Landau result stipulates the rapidity distribution to be
\cite{Lan53,Bel56}
\begin{eqnarray}
\label{Lan}
dN/d\lambda({\rm Landau}) \propto \exp \{ \sqrt{L^2-\lambda^2}\},
\end{eqnarray}
where the symbol $\lambda$ is often taken to be the rapidity variable
$y$ \cite{Mur04,Ste05,Ste07,Car73}.  In the original work of Landau
and his collaborator in \cite{Lan53,Bel56}, the variable $\lambda$ is
used to represent the polar angle $\theta$ as $e^{-\lambda}=\theta$;
there is the question whether the variable $\lambda$ in the Landau
rapidity distribution (\ref{Lan}) should be taken as the rapidity
variable $y$ \cite{Mur04,Ste05,Ste07,Car73} or the pseudorapdity
variable $\eta$ \cite{Sar06} appropriate to describe the polar angle.
Such a distinction between the rapidity and pseudorapidity variables
is quantitatively important because the shape of the distributions in
these two variables are different near the region of small rapidities
\cite{Won94}.  Secondly, the Gaussian rapidity distribution
(\ref{gau}) used in the analyses of Refs.\ \cite{Mur04,Ste05,Ste07},
as well as in the earlier work of \cite{Car73}, is only an
approximation of the original Landau distribution of Eq.\ (\ref{Lan}).
The Gaussian approximation is valid only for $\lambda \ll L$, when we
expand the square-root function in powers of $\lambda/L$.  There is
the question whether it is appropriate to apply the Gaussian
distribution (\ref{gau}) to the whole range of $\lambda$, including
the region of $\lambda$ for which the expansion condition $\lambda \ll
L$ is not satisfied.  The Landau distribution (\ref{Lan}) and the
Gaussian distribution (\ref{gau}) are in fact different distributions.
While the original Landau distribution may be considered to receive
theoretical support in Landau hydrodynamics as justified in
Refs. \cite{Lan53,Bel56}, a firm theoretical foundation for the
Gaussian distribution (\ref{gau}) in Landau hydrodynamics is still
lacking.  Finally, if one does not make the expansion of the
square-root function and one keeps the original Landau distribution of
Eq.\ (\ref{Lan}), then there is the quantitative question \cite{Cha74}
whether this original Landau distribution with the theoretical $L$
value will give results that agree with experimental data.

In view of the above unanswered questions, our task in reviewing the
Landau hydrodynamical model will need to ensure that we are dealing
with the rapidity variable $y$ and not the pseudorapidity variable.
We need to be careful about various numerical factors so as to obtain
a quantitative determination of the parameters in the final theoretical
results.  Finally, we need to ascertain whether the theoretical
results agree with experimental data.  If we succeed in resolving
these questions, we will pave the way for its future application to
other problems in high-energy heavy-ion collisions.

\vspace*{0.5cm} \section {\bf Total Number of Produced Particles }
\vspace{0.3cm}

Before we study the dynamics in detail, we would like to examine the
total number of particles produced in a central heavy-ion collision in
Landau hydrodynamics. Landau assumed that the hydrodynamical motion of
the fluid after the initial collision process is adiabatic.  He argued
that the only thing that can destroy adiabaticity would be the shock
wave.  They occur at the initial compressional stage of the collision
process \cite{Won74}.  It is hard to imagine how they could be formed
in the bulk part during the subsequent expansion phase after the
initial compression and thermalization.  Landau therefore assumed that
during the longitudinal and transverse expansion phase under
consideration, the entropy content of the the individual region
remains unchanged.  The total entropy of the system is therefore
unchanged and can be evaluated at the initial stage of the overlapped
and compressed system.

The total entropy content is a very useful quantity because it is
closely related to the total particle number.  From the consideration
of the thermodynamical properties of many elementary systems, Landau
found that the ratio of the entropy density to the number density for
a thermally equilibrated system is nearly a constant within the
temperature region of interest.  Landau therefore postulated that the
number density is proportional to the entropy density. Thus, by
collecting all fluid elements, the total number of particles is
proportional to the total entropy.  As the total entropy of the system
is unchanged during the hydrodynamical evolution, the total number of
observed particles can be determined from the initial entropy of the
system.

We work in the center-of-mass system and consider the central
collision of two equal nuclei, each of mass number $A$, at a
nucleon-nucleon center-of-mass energy $\sqrt{s_{_{NN}}}$.  Consider
first the case of central AA collisions with $A \gg 1$ such that
nucleons of one nucleus collide with a large numbers of nucleons of
the other nucleus and the whole energy content is used in particle
production.  This is the case of ``full stopping''.  The total energy
content of the system is
\begin{eqnarray}
E=\sqrt{s_{_{NN}}}A.
\end{eqnarray}
The initial compressed system is contained in a volume that is Lorentz
contracted to
\begin{eqnarray}
V=\frac{4\pi}{3} (a/2) ^3/\gamma ,
\end{eqnarray}
where the nuclear radius $a/2$ is related to the mass number by
\begin{eqnarray}
a/2=r_0 A^{1/3},
\end{eqnarray}
and $r_0=1.2$ fm.  The energy density of the system is therefore
\begin{eqnarray}
\epsilon = E/V= \gamma \sqrt{s_{_{NN}}}/(4\pi r_0^3/3),
\end{eqnarray}
which is independent of $A$ and depends on energy as $s_{_{NN}}$.
For a system in local thermal equilibrium, the entropy density
$\sigma$ is related to the energy density by
\begin{eqnarray}
\sigma = {\rm constant~}\epsilon^{3/4}.
\end{eqnarray}
The total entropy content of the system is therefore
\begin{eqnarray}
S=\sigma V ={\rm ~constant~} s_{_{NN}}^{1/4} A.  
\end{eqnarray}
With Landau's assumption relating entropy and particle number, 
$N\propto S$, the total number of particles produced is
\begin{eqnarray}
N \propto s_{_{NN}}^{1/4} A,  
\end{eqnarray}
and the total number of produced charged particles per participant
pair is
\begin{eqnarray}
\label{ncha}
N_{\rm ch}/A = 
N_{\rm ch}/(N_{\rm part}/2) = K (\sqrt{s_{_{NN}}}/{\rm GeV})^{1/2}, 
\end{eqnarray}
where $K$ can be determined phenomenologically by comparison with
experimental data.

\begin{figure} [h]
\includegraphics[angle=0,scale=0.50]{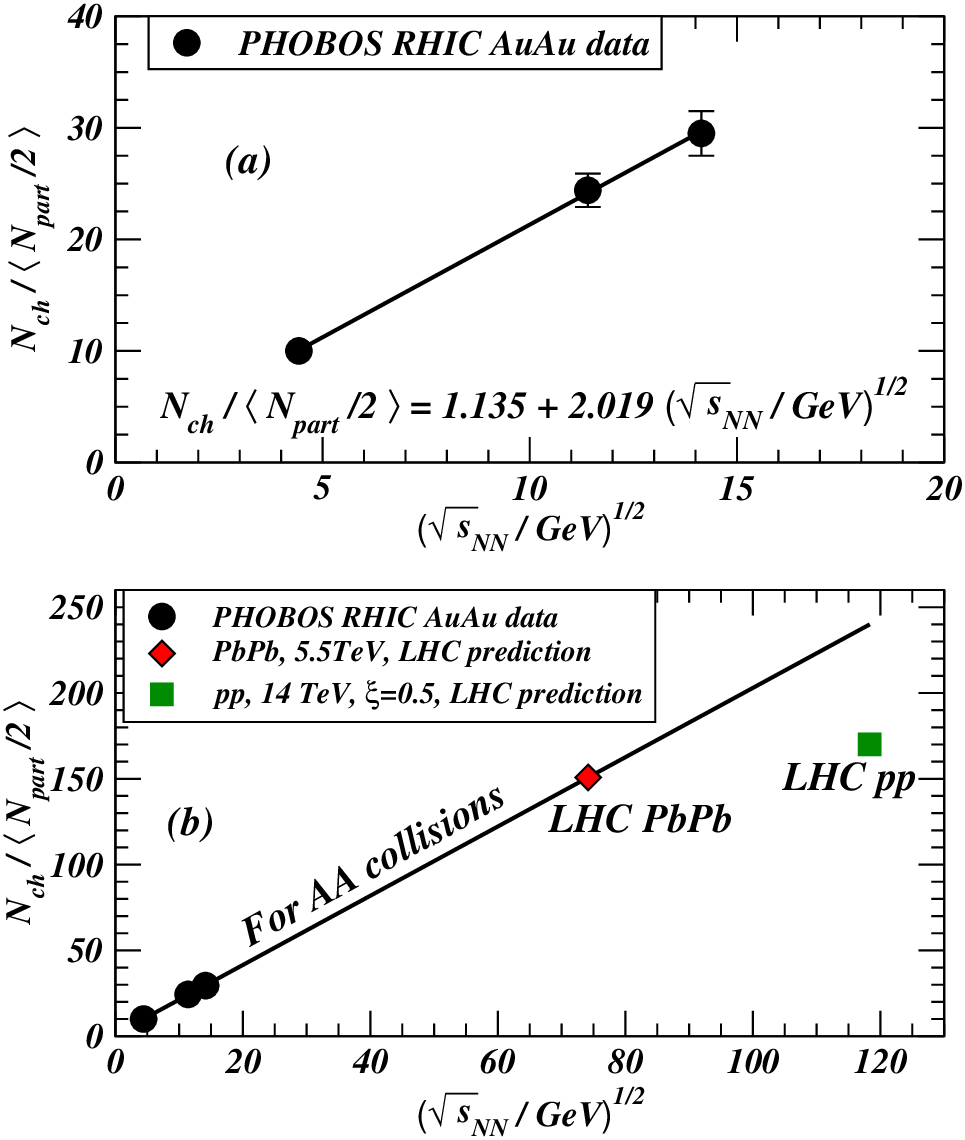}
\vspace*{0.0cm} 
\caption{Total number of produced charged particles per pair of
participants, $N_{\rm ch}/(N_{\rm part}/2)$, as a function of
$(\sqrt{s_{_{NN}}}/{\rm GeV})^{1/2}$.  (a) PHOBOS data $N_{\rm
ch}/(N_{\rm part}/2)$ data for central AuAu collisions at different
$(\sqrt{s_{_{NN}}}/{\rm GeV})^{1/2}$ and the Landau
hydrodynamical model fit, and (b) the extrapolation of the Landau
hydrodynamical model to LHC energies.}
\end{figure}

In Fig.\ 1(a), we show the PHOBOS data of $N_{\rm ch}/(N_{\rm
  part}/2)$ as a function of $(\sqrt{s_{{NN}}}/{\rm GeV})^{1/2}$ for
  central AuAu collisions in RHIC \cite{Ste05,Ste07,Bus04}.  The RHIC
  AuAu data can be parametrized as
\begin{eqnarray}
\label{system}
N_{\rm ch}/(N_{\rm part}/2) = 1.135 + 2.019 
(\sqrt{s_{_{NN}}}/{\rm GeV})^{1/2},
\end{eqnarray}
where the constant 1.135 arise from the leading nucleons.  The
constant $K$ as determined from the data is $K= 2.018$ which agrees
with the earlier estimate of $K=2$ \cite{Lan53,Bel56}.  

Consider next pp and p\= p collisions in which not all the energy of
$\sqrt{s_{_{NN}}}$ is used in particle production, as the leading
particles carry a substantial fraction of the initial energy.  If we
denote the particle production energy fraction in pp and p\= p
collisions by $\xi$, then Eq.\ (\ref{ncha}) is modified to be
\begin{eqnarray}
N_{\rm ch}= 
K (\xi \sqrt{ s_{_{NN}}}/{\rm GeV})^{1/2}. 
\end{eqnarray}
Comparison of the charged particle multiplicity in pp, p\= p, and
$e^+e^-$ collisions indicates that the particle production energy
fraction $\xi$ for pp and p\= p collisions is approximately 0.5
\cite{Bas81,Won94,Ste05,Ste07}.  In contrast, the case of RHIC AA data
in high-energy heavy-ion collisions corresponds to full nuclear
stopping with $\xi=1$ \cite{Ste05,Ste07}.

In Fig.\ 1(b), we show the predictions for the charge particle
multiplicity per pair of participants for collisions at LHC energies.
For pp collisions at 14 TeV with a particle production energy fraction
$\xi=0.5$, $N_{\rm ch}$ is predicted to be 170.  For central PbPb
collisions at $\sqrt{s_{_{NN}}}= 5.5$ TeV with full nuclear stopping
$(\xi=1)$, $N_{\rm ch}/(N_{\rm part}/2)$ is predicted to be 151.

\vspace*{0.5cm} \section {\bf Longitudinal Hydrodynamical Expansion }
\vspace{0.3cm}

We proceed to examine the dynamics of the longitudinal and transverse
expansions.  Among the coordinates $(t,z,x,y)\equiv
(x^0,x^1,x^2,x^3)$ used as the arena to describe the fluid, Landau
suggested a method to split the problem into two stages.  The first
stage consists of simultaneous and independent expansions along the
longitudinal and the transverse directions.  For the one-dimensional
longitudinal expansion, the equation of hydrodynamics is
\begin{eqnarray}
\label{T00}
\frac{\partial T^{00}}{\partial t} + \frac{\partial T^{01}}{\partial z} =0,\\
\label{T01}
\frac{\partial T^{01}}{\partial t} + \frac{\partial T^{11}}{\partial z} =0,
\end{eqnarray}
where
\begin{eqnarray}
T^{\mu \nu}=(\epsilon + p)u^\mu u^\nu - p g^{\mu \nu}.
\end{eqnarray}
In hydrodynamical calculations, we need an equation of state.  We
shall assume for simplicity the relativistic equation of state
\begin{eqnarray}
\label{eqst}
p=\epsilon/3,
\end{eqnarray}
which allows one to express the local pressure $p$ as a function of the
energy density $\epsilon$.
It is convenient to represent the velocity fields $(u^0, u^1)$ by the
flow rapidity $y$
\begin{subequations}
\begin{eqnarray}
\label{u0}
u^0&=& \cosh y, \\
\label{u1}
u^1&=& \sinh y. 
\end{eqnarray}
\end{subequations}
In this form, the property of the velocity fields,
$(u^0)^2-(u^1)^2=1$, is readily satisfied. 
A particle of mass $m$ flowing with such a
rapidity $y$ will have a momentum $(p^0,p^1)=(mu^0,mu^1)=(m\cosh
y,m\sinh y)$.  

It is convenient to introduce the light-cone coordinates $t_+$ and $t_-$
\begin{subequations}
\begin{eqnarray}
t_+&=& t+z, \\
t_-&=& t-z, 
\end{eqnarray}
\end{subequations}
and their logarithmic representations  $(y_+,y_-)$ defined by
\begin{eqnarray}
\label{ypm}
y_{\pm} = \ln \{ t_{\pm} /\Delta \} = \ln \{ (t {\pm} z) /\Delta \}.
\end{eqnarray}
The advantage of using the these variables is that the hydrodynamical
equations and the corresponding solutions are quite simple in these
variables.  The knowledge of the fluid energy density $\epsilon$ and
the flow rapidity $y$, as a function of $(t_+, t_-)$ or $(y_+,y_-)$
will provide a complete hydrodynamical description of the whole
system.

Exercise (1.1) shows that the hydrodynamical equations (\ref{T00}) and
(\ref{T01}) become
\begin{subequations}
\begin{eqnarray}
\label{eq1da}
\frac{\partial \epsilon}{\partial t_+} 
+ 2 \frac{\partial (\epsilon e^{-2y}) }{\partial t_-}&=&0,\\
\label{eq1db}
2 \frac{\partial (\epsilon e^{2y}) }{\partial t_+} 
+  \frac{\partial \epsilon  }{\partial t_-}&=&0.
\end{eqnarray}
\end{subequations}
The $t_+$ and $t_-$ coordinates are light-like because they lie on the
light cone.  If we view the $t_+$ coordinate as approaching time-like
and the $t_-$ as approaching space-like, then the first equation can
be considered the equation of continuity and the second the Euler
equation of the hydrodynamical current.  Conversely, if we view the
$t_-$ coordinate as approaching time-like and the $t_+$ as approaching
space-like, then the second equation can be considered the equation of
continuity and the first the Euler equation of the hydrodynamical
current.

For the first stage of one-dimensional hydrodynamics, the exact
solution for an initially uniform slab has been obtained \cite{Kha54}
and discussed in
\cite{Bel56,Kha54,Bel56a,Ama57,Sri92,Ham04,Pra07,Bia07,Cso07,Cso08,Beu08}.
There is in addition an approximate particular solution that is very
simple and applicable to the bulk part of the fluid
\cite{Lan53,Bel56}.  In view of the matching of the solution to an
approximate three-dimensional motion in the second stage, it suffices
to consider the approximate solution of longitudinal hydrodynamical
expansion in the present discussions.

The approximate  particular solution of the hydrodynamical equation is
\cite{Bel56}
\begin{subequations}
\begin{eqnarray}
\label{sole}
\epsilon (y_+,y_-) & = &  \epsilon_0\exp\left \{ -\frac{4}{3}(y_+ + y_- -
\sqrt{y_+y_-}) \right \},\\
\label{soly}
y (y_+,y_-) & = &  (y_+- y_-)/2.
\end{eqnarray}
\end{subequations}
The rapidity solution can also be written alternatively as
\begin{eqnarray}
\label{sol2}
e^{ 2 y(y_+,y_-) } = \frac{t_+}{t_-} = \frac{t + z}{t-z}.
\end{eqnarray}
The corresponding velocity field $u^0$ is related to the fluid
coordinates by
\begin{eqnarray}
\label{u0I}
u^0(y_+,y_-) = \frac{t}{\sqrt{t-z}\sqrt{t+z}} .
\end{eqnarray}
The constant $\epsilon_0$ in Eq.\ (\ref{sole}) is related to the
initial energy density at $(y_{+0}, y_{-0})$ by
\begin{eqnarray}
\epsilon_0=
\epsilon(y_{+0},y_{-0}) e^{\phi_0},
\end{eqnarray}
where $\phi_0$ is 
\begin{eqnarray}
\phi_0= \frac{4}{3} ( y_{+0} +y_{-0}
-\sqrt{y_{+0}y_{-0}}).
\end{eqnarray}
We can easily prove by direct substitution that (\ref{sole}) and
(\ref{soly}) (or (\ref{sol2})) are approximate particular solutions of
the hydrodynamical equations (\ref{eq1da}) and (\ref{eq1db}).  First,
substituting Eq.\ (\ref{sol2}) into the hydrodynamical equations, we
obtain
\begin{subequations}
\begin{eqnarray}
\frac{\partial \epsilon}{\partial t_+} 
+ 2 \left [ \frac{\partial \epsilon  }{\partial t_-}
+\frac{1}{t_-} \right ] \frac{t_-}{t_+}
&=&0,\\
2  \left [\frac{\partial \epsilon  }{\partial t_+} +\frac{1}{t_+} \right ]
 \frac{t_+}{t_-}
+  \frac{\partial \epsilon  }{\partial t_-}&=&0.
\end{eqnarray}
\end{subequations}
We write out $t_-/t_+$ in the second equation and substitute it into
the first equation, and we get
\begin{eqnarray}
\frac{\partial \epsilon}{\partial t_+} 
\frac{\partial \epsilon}{\partial t_-} 
- 4 \left [ \frac{\partial \epsilon  }{\partial t_-}
+\frac{1}{t_-} \right ] 
  \left [\frac{\partial \epsilon  }{\partial t_+} +\frac{1}{t_+} \right ]
=0
\end{eqnarray}
We multiply this expression by $t_+ t_-$ and change into the
logarithm variables $y_+$ and $y_-$, then the above equation becomes
\begin{eqnarray}
\frac{\partial \epsilon}{\partial y_+} 
\frac{\partial \epsilon}{\partial y_-} 
- 4 \left [ \frac{\partial \epsilon  }{\partial y_-}
+1\right ] 
  \left [\frac{\partial \epsilon  }{\partial y_+} +1 \right ]
=0
\end{eqnarray}
If we now substitute Eq.\ (\ref{sole}) for $\epsilon$ into the
lefthand side of the above equation, we find that the lefthand side
gives zero, indicating that Eqs.\ (\ref{sole}) and (\ref{soly}) are
indeed the approximate  particular solution of the hydrodynamical equation.

The above solutions (\ref{sole}) and (\ref{soly}) are applicable for
both positive and negative $z$ values and contains the proper
reflectional symmetry with respect to the interchange of $y_+$ and
$y_-$.  The solutions are appropriate for the description of the
dynamics of the dense hot matter in the first stage of
longitudinal expansion.  For simplicity, we shall concentrate our
attention on one of the two symmetric sides, the side with positive
$z$.  

The simple approximate  solution of (\ref{sole}) and (\ref{soly}) have
limitations.  Because of the nature of the approximate  solution, it cannot
describe the boundary layer for which $t-z <\Delta$ and $y_-$ becomes
negative.  At the boundary layer, thermodynamical quantities such as
the energy density and the entropy density decrease rapidly and it is
not included into the region for the applications of the solution of
Eqs.\ (\ref{sole}).  In highly relativistic collisions $\Delta/a$ is
given by $(2m_p /\sqrt{s_{_{NN}}})$, which is a small quantity.  The
region excluded from the approximate solution is not significant in a
general description of the fluid.

There is another minor disadvantage of the solution.  Because of the
nature of the solution for the bulk matter, the solution in Eq.\
(\ref{sole}) provides only limited choice on the initial conditions,
within the form as specified by the simple functions in these
equations.  However, the dynamics of the system is likely to be
insensitive to the fine characteristics of the initial density
distribution of the contracted disk of thickness $\Delta$ along the
longitudinal direction.  A thin slab of matter of the right dimensions
within the Landau model will likely capture the dominant features of
the evolution dynamics.

\begin{figure} [h]
\includegraphics[angle=0,scale=0.50]{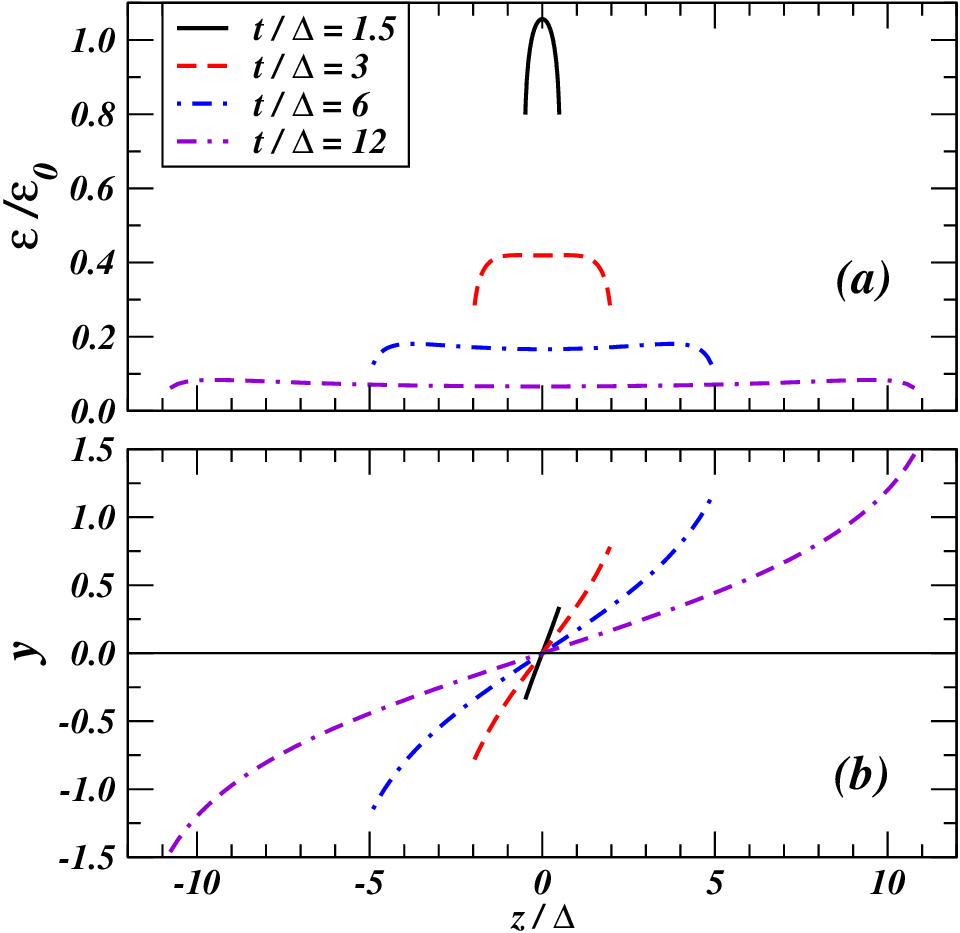}
\vspace*{0.0cm} 
\caption{ Approximate Landau hydrodynamical solution of the
  one-dimensional longitudinal expansion of the system with initial
  energy density $\epsilon_0$, as a function of the longitudinal
  coordinate $z$ and various times $t$ in unit of the slab thickness
  $\Delta$. }
\end{figure}

The solutions $\epsilon (t_+,t_-)$ and $y(t_+,t_-)$ can be transformed
into $\epsilon (t,z)$ and $y (t,z)$ by a simple change of variables.
We show the solution $\epsilon (t,z)$ and $y(t,z)$ as a function of
$z/\Delta$ for different $t$ in Fig.\ 2(a) and 2(b) respectively.  As
one observes, the maximum energy density decreases nearly inversely as
a function of $t$.  One notes that the energy density in the central
region decreases in time approximately as $1/t$, as the matter expands
outward.  The flow rapidity is initially close to zero.  As the bulk
matter expands longitudinally, the magnitude of the rapidity field
increases at large values of $|z|/\Delta$.  We show the velocity field
$u^0(t,z)$ and the corresponding longitudinal velocity $v_z (t,z)$ as
a function of $z/\Delta$ for different $t$ in Fig.\ 3(a) and 3(b).
The magnitude of the longitudinal velocity of the fluid is large at
the region of $|z|$ close to $t$.

\begin{figure} [h]
\includegraphics[angle=0,scale=0.50]{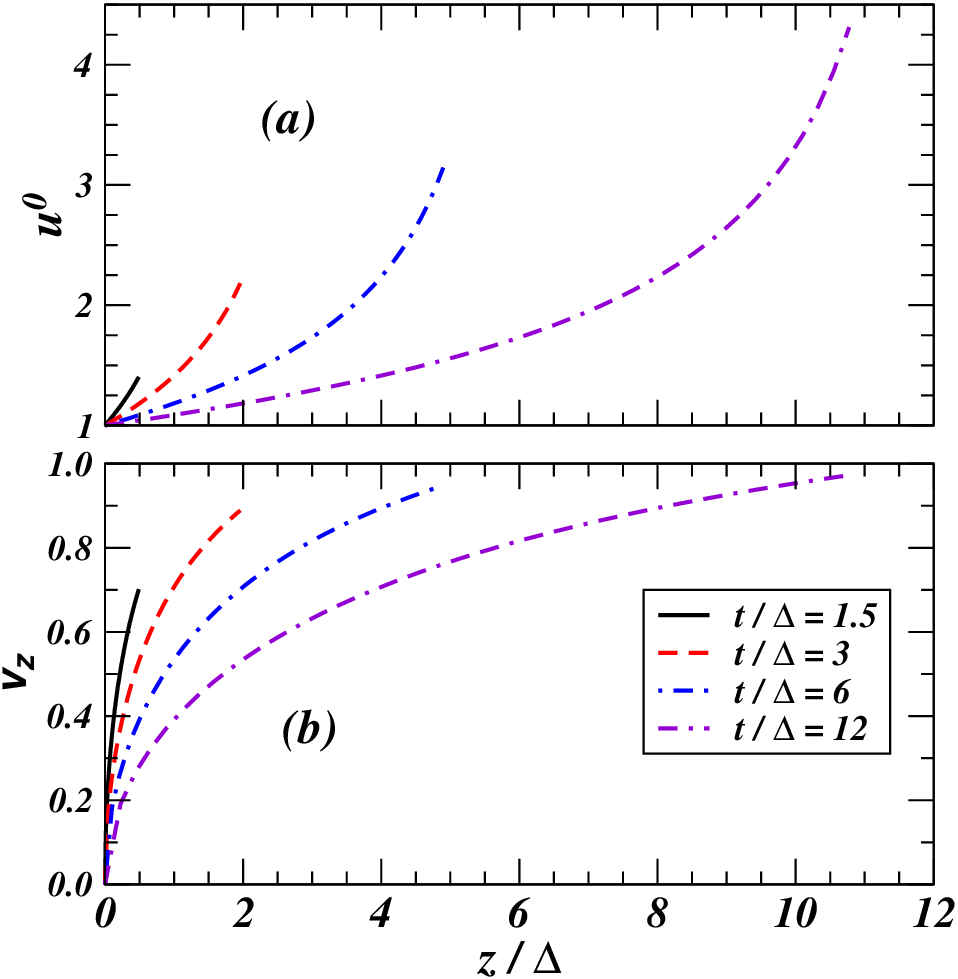}
\vspace*{0.0cm} 
\caption{ The proper velocity $u^0$ and the velocity $v_z$ along the
  longitudinal $z$-direction as a function of $z$ and $t$ Landau
  hydrodynamics. }
\end{figure}

The hydrodynamical solution Eq. (\ref{sole}) can be equivalently
expressed as a function of any other two independent variables.  In
our problem, it is desirable to use the variables $t$ and $y$ because
the condition for the transverse displacement to exceed the initial
transverse dimension is most conveniently written for a fixed time $t$,
and the dynamics after the transversality condition is satisfied leads
to the freezing of the rapidity $y$.

In terms of the $(t,y)$ variables, we can express $y_\pm$ as
\begin{eqnarray}
y_\pm=\ln\left ( {t}/{\cosh y \, \Delta } \right ) \pm y.
\end{eqnarray}
Therefore the hydrodynamical solution Eq.\ (\ref{sole}) can be
re-written in terms of $(t,y)$ as
\begin{subequations}
\begin{eqnarray}
\epsilon (t,y) & =& \epsilon_0\exp\left \{ -\frac{4}{3} (2
\ln\left ( {t}/{\cosh y \, \Delta } \right ) - \sqrt{ [\ln\left (
    {t}/{\cosh y\, \Delta } \right )]^2-y^2 }) \right \},\\
z(t,y) & =& t\, {\sinh y}/{\cosh y}.
\end{eqnarray}
\end{subequations}
We can further express the solution as a function of $(\tau,y)$ to
compare Landau hydrodynamics with Bjorken hydrodynamics.  We make the
transformation
\begin{subequations}
\begin{eqnarray}
t&=& \tau \cosh y, \\
z&=& \tau \sinh y. 
\end{eqnarray}
\end{subequations}
We then have
\begin{eqnarray}
y_\pm=\ln (\tau/\Delta) \pm y.
\end{eqnarray}
The energy density in Landau hydrodynamics, expressed in terms of 
$\tau$ and $y$ is then
\begin{eqnarray}
\epsilon (\tau, y)  =  \epsilon_0\exp\left \{ 
-\frac{4}{3}\left [ 2\ln({\tau}/{\Delta})-\sqrt{[\ln (\tau/\Delta)]^2-y^2}
\right ] \right \}.
\end{eqnarray}
In the  region $y\ll \ln (\tau/\Delta)$, we have
\begin{eqnarray}
\epsilon (\tau, y)  =  \epsilon_0\exp\left \{
-\frac{4}{3}\ln({\tau}/{\Delta}) \right \} \propto \frac{1}{\tau^{4/3}},
\end{eqnarray}
which is the Bjorken hydrodynamics results.  Therefore, in the region
of small rapidities with $|y| \ll\ln (\tau/\Delta)$, Landau
hydrodynamics and Bjorken hydrodynamics coincide. However, in other
regions when $|y|$ is not very smaller in comparison with $\ln
(\tau/\Delta)$, the two types of hydrodynamics differ.  In general,
because Landau hydrodynamics covers a wider range of rapidities which
may not be small, it is a more realistic description for the evolution
of the hydrodynamical system.

\vskip 0.5cm \small
\baselineskip=12pt
\noindent
$\bigoplus [\![$ Exercise 1.1  Show that the
hydrodynamical equations for $\epsilon$ and $y$ are given by
Eqs. (\ref{eq1da}) and (\ref{eq1db}).

\noindent
The energy momentum tensor $T^{00}$ is equal to
\begin{eqnarray}
T^{00}&=&(\epsilon +p) u^0 u^0 - p
\nonumber \\
&=& (\epsilon +p) \cosh^2 y -p.
\end{eqnarray}     
Upon using the equation of state $p=\epsilon/3$ in Eq.\ (\ref{eqst}),
the energy momentum tensor can be expressed in terms of $\epsilon$ and
$y$.  
\begin{eqnarray}
T^{00}= \epsilon (\cosh^2 y + \frac {1}{3}\sinh y).
\end{eqnarray}     
We have also,
\begin{eqnarray}
T^{01}&=&(\epsilon +p) u^0 u^1 
\nonumber \\
&=& (4/3)\epsilon  \cosh y \sinh y,
\end{eqnarray}     
\begin{eqnarray}
T^{11}&=&(\epsilon +p) u^1 u^1 + p
\nonumber \\
&=& \epsilon (\sinh^2 y + \frac {1}{3} \cosh^2 y).
\end{eqnarray}     
The transformation of $(t,z)$ to $(t_+, t_-)$ lead to the following
relation between the derivatives,
\begin{eqnarray}
 \frac{\partial}{\partial t}=
 \frac {\partial t_+} {\partial t}
 \frac {\partial }    {\partial t_+}
+\frac {\partial t_-} {\partial t}
 \frac {\partial }    {\partial t_-}
= \frac {\partial }    {\partial t_+}
+\frac {\partial }    {\partial t_-}.
\end{eqnarray}
\begin{eqnarray}
 \frac{\partial}{\partial z}=
 \frac {\partial t_+} {\partial z}
 \frac {\partial }    {\partial t_+}
+\frac {\partial t_-} {\partial z}
 \frac {\partial }    {\partial t_-}
= \frac {\partial }    {\partial t_+}
-\frac {\partial }    {\partial t_-}.
\end{eqnarray}
Therefore, Eqs. (\ref{T00}) and (\ref{T01}) become
\begin{eqnarray}
\frac{dT^{00}}{dt} + \frac{dT^{01}}{dz} =
\frac{\partial}{\partial t_+}
\left [\frac{4}{3} \epsilon \sinh y \cosh y +
 \epsilon (\cosh^2 y+\frac {1}{3}  \sinh^2 y)\right ]
+\frac{\partial}{\partial t_-}
\left [-\frac{4}{3} \epsilon \sinh y \cosh y +
 \epsilon (\cosh^2 y+\frac {1}{3}  \sinh^2 y)\right ].
\end{eqnarray}
\begin{eqnarray}
\frac{dT^{01}}{dt} + \frac{dT^{11}}{dz} =
\frac{\partial}{\partial t_+}
\left [\frac{4}{3} \epsilon \sinh y \cosh y +
 \epsilon (\sinh^2 y+\frac {1}{3}  \cosh^2 y)\right ]
+\frac{\partial}{\partial t_-}
\left [\frac{4}{3} \epsilon \sinh y \cosh y -
 \epsilon (\sinh^2 y+\frac {1}{3}  \cosh^2 y)\right ].
\end{eqnarray}
By forming the difference of the above two
equations, we get
\begin{eqnarray}
\frac{\partial}{\partial t_+}
[ \epsilon (\cosh^2y + \frac{1}{3} \sinh^2 y)]
+\frac{\partial}{\partial t_-}
[-\frac{8}{3} \epsilon \sinh y \cosh y 
+\frac{4}{3} \epsilon (\cosh^2 y + \sinh^2 y)]
=0.
\end{eqnarray}
This can be simplified to be
\begin{eqnarray}
\frac{\partial \epsilon}{\partial t_+}
+2\frac{\partial (\epsilon e^{-2y}) }{\partial t_-}
=0,
\end{eqnarray}
which is Eq. (\ref{eq1da}).  Similarly, by forming the sum of the two
equations, we get
\begin{eqnarray}
\frac{\partial}{\partial t_+}
\left [ \frac{4}{3} \epsilon (\cosh^2y + \sinh^2)
 +\frac{8}{3} \epsilon  \sinh y \cosh y \right ]
+\frac{\partial}{\partial t_-}
\left [\frac{2}{3} \epsilon (\cosh^2 y- \sinh^2 y \right ]
=0.
\end{eqnarray}
This leads to
\begin{eqnarray}
2 \frac{\partial (\epsilon e^{2y}) }{\partial t_+} 
+  \frac{\partial \epsilon  }{\partial t_-}&=&0,
\end{eqnarray}
which is Eq. (\ref{eq1db}).\hfill $]\!]\bigoplus$

\large
\vspace*{0.5cm}
\section {\bf 
Transverse Expansion }
\vspace{0.3cm} 

The initial configuration is much thinner in the longitudinal
direction than in the transverse directions. Therefore, in the first
stage of evolution during the fast one-dimensional longitudinal
expansion, there is a simultaneous but slower transverse expansion.
The difference in the expansion speeds allows Landau to treat the
longitudinal and transverse dynamics as independent expansions.  The
rate of transverse expansion can then be obtained to provide an
approximate description of the dynamics of the system.

We shall consider first the case of a central collision, for which
$a_y=a_x=a$.  The case of non-central collisions will be discussed in
Section IX.  The transverse expansion is governed by the Euler equation
along one of the transverse directions, which can be taken to be along
the $x$ direction,
\begin{eqnarray}
\label{T02}
\frac{\partial T^{02}}{\partial t} 
+ \frac{\partial T^{22}}{\partial x} =0,
\end{eqnarray}
where 
\begin{eqnarray}
T^{02}=(\epsilon+p) u^0 u^2=\frac{4}{3} \epsilon (u^0)^2 v_x,
\end{eqnarray}
and we have used the relation $u^2=u^0 v_x$.  The energy-momentum
tensor $T^{22}$ is
\begin{eqnarray}
T^{22}=(\epsilon+p) (u^2)^2 -p g^{22}=\frac{4}{3} \epsilon (u^0)^2
(v_x)^2+p.
\end{eqnarray}
As the transverse expansion is relatively slow, we can neglect the
first term on the righthand side of the above expression and keep only
the pressure term $p$.

In Landau's method of splitting the equations, one makes the
approximation that during the first stage the quantities $\epsilon$
and $y$ as a function of $t$ and $z$ have been independently
determined in the one-dimensional longitudinal motion.  It suffices to
use Eq.\ (\ref{T02}) to determine the transverse displacement as a
function of time.  Equation\ (\ref{T02}) can therefore be approximated
as
\begin{eqnarray}
\label{trans}
\frac{4}{3} \epsilon (u^0)^2 \frac{\partial v_x}{\partial t}
=-\frac{\partial p}{\partial x}.
\end{eqnarray}
The transverse displacement $ x(t)$ (relative to zero displacement) as
a function of time $t$ is related to the acceleration $\partial
v_z/\partial t$ by
\begin{eqnarray}
 x (t) = 
 \frac{1}{2}\left ( \frac{\partial v_x}{\partial t} \right ) t^2. 
\end{eqnarray}
The pressure is $p=\epsilon/3$ at the center of the transverse region and
is zero at the radial surface $a/2$.  Therefore the equation for the
displacement is given from Eq.\ (\ref{trans}) by
\begin{eqnarray}
\label{displ}
\frac{4}{3} \epsilon (u^0)^2 \frac{ 2 x(t) }{ t^2}
= \frac{ \epsilon}{ 3a/2}. 
\end{eqnarray}
We note that there is a factor of 4 arising from the ratio of
$4\epsilon /3$ from $(\epsilon+p)$ on the lefthand side relative to
$\epsilon/3$ from the pressure $p$ on the righthand side.  However, in
the original formulation of Landau \cite{Lan53,Bel56}, this factor of
4 is taken to be unity for an order of magnitude estimate of the
transverse displacement.  For our purpose of making quantitative
comparison with experimental data, this factor of 4 cannot be
neglected. We shall find in Section 5 and 6 that the presence of this
factor of 4 leads to the modified rapidity distribution (\ref{new})
that agrees better with experimental $dN/dy$ data than the Landau
distribution of Eq.\ (\ref{Lan}).

From Eq.\ (\ref{displ}), the transverse displacement $x(t)$ during
the one-dimensional longitudinal expansion increases quadratically
with $t$ as
\begin{eqnarray}
\label{xt}
 x(t) = \frac{t^2}{4 a (u^0)^2 }=\frac{t^2}{4 a \cosh^2 y} .
\end{eqnarray}
For a fixed $t$, the transverse displacement $x(t)$ as a function of
the rapidity is greatest at $y=0$ and it decreases as $|y|$ increases.
The variation of the transverse displacement as a function of rapidity
leads to different times when the fluid element enters from the first
stage to the second stage of frozen rapidities.

\large
\vspace*{0.5cm}

\section {\bf 
Second Stage of Conic Flight}
\vspace{0.3cm} 

The description of one-dimensional solution of longitudinal expansion
with the accompanying transverse expansion in the first stage is
applicable as long as the transverse displacement $ x(t)$ is
sufficiently small compared with the transverse dimension $a$.  It
ceases to be applicable at the time $t_m$ when the transverse
displacement $ x(t_m)$ is equal to the transverse dimension $a$.
Landau suggested that when $ x(t_m)$ is equal to $a$ we need to switch
to a new type of solution in the second stage of fluid dynamics.

What type of fluid motion is expected after the fluid element enters
the second stage?  For a fluid element at $t\ge t_m$, the transverse
displacement $x(t)$ already has exceeded the initial transverse
dimension.  With the fluid element beyond the initial transverse
dimension, the derivatives of hydrodynamical quantities with respect
to both the transverse coordinates and $t_-$ are small.  Because of
the smallness of these derivatives, Landau argued that the
hydrodynamical forces become so small that they can be neglected in
the hydrodynamical equations at these locations.  The flow rapidity
$y$ will be frozen for $t\ge t_m$.  Freezing the rapidity of a fluid
element as a function of time is equivalent to freezing the opening
polar angle $\theta$ between the fluid trajectory and the collision
axis at these longitudinal locations.  The motion of the fluid element
with a fixed polar angle can be described as a `three-dimensional'
conic flight, with no change of the flow rapidity.

In mathematical terms, Landau's condition for rapidity freezeout
occurs at the time $t_m(y) $ when the transverse displacement satisfies
\cite{Lan53,Bel56}
\begin{eqnarray}
\label{xtt}
 x(t_m) = a,
\end{eqnarray}
which, as determined from Eqs.\ (\ref{xt}) and (\ref{xtt}), takes place at 
\begin{eqnarray}
\label{tmy}
t_m(y) = 2 a u^0 =2 a \cosh y. 
\end{eqnarray}
The switching time $t_m(y)$ increases with increasing magnitude of the
rapidity.

\begin{figure} [h]
\includegraphics[angle=0,scale=0.40]{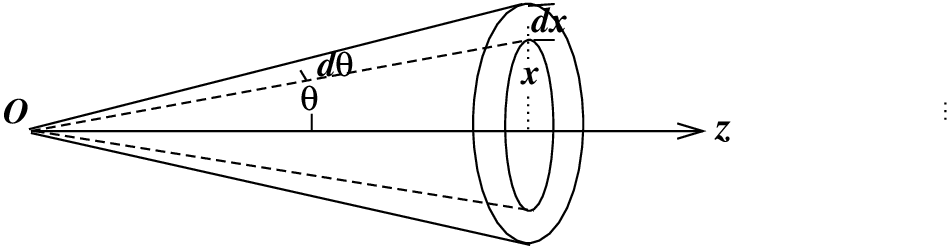}
\vspace*{0.0cm} 
\caption{ Schematic picture of the opening polar angle $\theta$
  between the fluid element trajectory and the longitudinal axis.  The
  conic energy and entropy within the angle element $d\theta$ is
  preserved in this second stage of conic flight, }
\end{figure}

The solution of the hydrodynamical variables during this second stage
of conic flight can be easily obtained.  In a conic flight with a cone
of opening polar angle $\theta$ within an angle element $d\theta$ as
shown in Fig.\ 5, the energy-momentum tensor and the entropy flux
within the cone element must be conserved as a function of time. The
cross sectional area of such a cone element is $2\pi x dx$.   So the
conservation of energy and entropy conic flow correspond to
\begin{eqnarray}
\label{de}
dE=\epsilon u^0 u^0 2\pi x dx = {\rm ~constant},
\end{eqnarray}
and
\begin{eqnarray}
dS=\sigma u^0 2\pi x dx = \epsilon^{3/4} u^0 2\pi x dx = {\rm ~constant},
\end{eqnarray}
where $\sigma$ is the entropy density.  Dividing the first equation by
the second equation, we get
\begin{eqnarray}
\epsilon^{1/4} u^0= {\rm ~constant},
\end{eqnarray}
which gives
\begin{eqnarray}
\label{ett}
\epsilon \propto \frac{1}{(u^0)^4}.
\end{eqnarray}
On the other hand, in the conic flight, $x$ and $dx$ are proportional
to $t$. Hence, Eq.\ (\ref{de}) gives
\begin{eqnarray}
\label{uutt}
\epsilon u^0u^0 t^2 = {\rm ~constant}.
\end{eqnarray}
Eqs.\ (\ref{ett}) and (\ref{uutt}) yield the dependence of various
quantities as a function of $t$,
\begin{eqnarray}
\epsilon \propto \frac{1}{t^4},
~~~\sigma \propto \frac{1}{t^3}, {\rm ~~~and~~~~}
u^0 \propto t.
\end{eqnarray}
These equations give the solution of the evolution of the fluid
elements as a function of time in the second stage.  By matching the
solutions at $t=t_m(y)$, the energy density and velocity fields at the
second stage for $t > t_m(y)$ is
\begin{subequations}
\begin{eqnarray}
\epsilon (t,y)& =& 
\epsilon (t_m,y)\, {t_m^4}/{t^4}\\
u^0(t,y)& = & u^0(t_m,y)\, {t}/{t_m}.
\end{eqnarray}
\end{subequations}
The velocity component $u^0$ of a fluid element increases linearly
with time in this second stage. In contrast, in the first stage, the
velocity component $u^0$ of a fluid element with fixed $t-z$ and $z$
comparable to $t$ increase with time only approximately as $t^{1/2}$,
as indicated by Eq.\ (\ref{u0I}).  Thus, the fluid element approaches
the speed of light faster in the second stage than in the first stage.

\large
\vspace*{0.5cm} 
\section{\bf Rapidity Distribution
in High Energy Heavy-Ion Collisions }
\vspace{0.3cm}

The picture that emerges from Landau hydrodynamics can be summarized
as follows.  For an initial configuration of a thin disk of dense
matter at a high temperature and pressure, the first stage of the
motion is a one-dimensional longitudinal expansion with a simultaneous
transverse expansion.  The transverse expansion lead to a transverse
displacement.  When the magnitude of the transverse displacement
exceeds the initial transverse dimension, forces acting on the fluid
element becomes small and the fluid elements will proceed to the
second stage of conic flight with a frozen rapidity.  As the
transverse displacement depends on rapidity, and the transverse
displacement magnitude increases with increasing rapidity magnitude,
the moment when the fluid element switches from the first stage to the
second stage depends on the rapidity.  The final rapidity distribution
of particles is therefore given by the rapidity distribution of the
particles at the matching time $t_m(y) $.

We shall first evaluate the entropy distribution as a function of
rapidity $y$ and time $t$ in the first stage of hydrodynamics.
Consider a slab element $dz$ at $z$ at a fixed time $t$.  The entropy
within the slab element is
\begin{eqnarray}
dS=\sigma u^0 dz.
\end{eqnarray}
Using the solution (\ref{sol2}), we can express $z$ as a function of
$t$ and rapidity $y$ during the one-dimensional longitudinal
expansion,
\begin{eqnarray}
\label{zt}
z=t\, {\sinh y}/{\cosh y}.
\end{eqnarray}
For a fixed value of $t$, we therefore obtain
\begin{eqnarray}
dS &=& \sigma u^0 t dy/ {\cosh^2 y} 
\nonumber\\
&=&  {\sigma t dy}/{\cosh y} .
\end{eqnarray}
The entropy density $\sigma$ is related to $\epsilon$ by $\sigma =c
\epsilon^{3/4}$ and $\epsilon$ is given by (\ref{sole}).  We obtain
the entropy element at the time $t$,
\begin{eqnarray}
dS &=& c \epsilon_0^{3/4} \exp\{ -(y_++y_- 
-\sqrt{y_+y_-} )\} \, {t ~dy}/{\cosh y}. 
\end{eqnarray}
In the second stage, different fluid elements with different
rapidities switch to conic flight at different time $t_m(y)$.  The
rapidity is frozen after $t> t_m(y) $.  The entropy element after
freezeout needs to be evaluated at the switching time $t=t_m(y) $
\begin{eqnarray}
\label{rap}
dS &=& c \epsilon_0^{3/4} \left [\exp\{ -(y_++y_- 
-\sqrt{y_+y_-}) \} \frac{t}{\cosh y} \right ]_{t=t_m(y) } dy
\end{eqnarray}
To evaluate the square-bracketed quantity at $t=t_m(y) $, we obtain from
Eq.\ (\ref{ypm}) and (\ref{zt}) that
\begin{eqnarray}
e^{y_\pm}=\frac{t}{\Delta} \frac{e^{\pm y}}{\cosh y}.
\end{eqnarray}
Therefore, we have
\begin{eqnarray}
\label{extm}
e^{y_\pm} \bigl | _{t=t_m(y) } =\frac{t_m(y) }{\Delta} \frac{e^{\pm y}}{\cosh y}
=\frac{2a}{\Delta}e^{\pm y} ,
\end{eqnarray}
which gives 
\begin{eqnarray}
\label{yyy}
y_{\pm} \bigl | _{t=t_m(y) }
= \ln \left ( {2a}/{\Delta}\right ) \pm y.
\end{eqnarray}
We note that 
\begin{eqnarray}
\ln \left ( {2a}/{\Delta} \right )=y_b=L+\ln 2,
\end{eqnarray}
where $y_b$ is the beam rapidity in the center-of-mass system,
\begin{eqnarray}
y_b = \cosh^{-1} \left ( {\sqrt{s_{NN}}}/{2m_p} \right )
\doteq \ln \left ( { \sqrt{s_{NN}}}/{m_p} \right ).
\end{eqnarray}
The entropy element of Eq.\ (\ref{rap}) is therefore
\begin{eqnarray}
dS &=& c 
\epsilon_0^{3/4} 2 a \exp\{ -2y_b 
+\sqrt{y_b^2-y^2} \}dy
\end{eqnarray}
As the entropy is proportional to the number of particles, we obtain
the rapidity distribution of the particle number 
\begin{eqnarray}
\label{new}
\frac{dN}{dy} &\propto& \exp\{\sqrt{y_b^2-y^2} \}.
\end{eqnarray}
which differs from Landau's rapidity distribution of Eq.\ (\ref{Lan}),
with the beam rapidity $y_b=L+\ln2$ replacing Landau's $L$.
Therefore, we find that the rapidity distribution in the
center-of-mass system should be modified from the original Landau
distribution.

While many steps of the formulation are the same, the main difference
between our formulation and Landau's appears to be the additional
factor of 2 in Eq.\ (\ref{extm}) and (\ref{tmy}) in the new
formulation.  This factor can be traced back to the factor of 4 in the
ratio of $4\epsilon/3$ from $(\epsilon+p)$ on the left hand side of
Eq. (\ref{displ}) and $\epsilon/3$ from the pressure $p$ on the right
hand side.  In Landau's formulation, this factor of 4 is taken to be
unity for an order-of-magnitude estimate of the transverse expansion.
Hence, there can be a modification of Landau's expression for the
rapidity distribution.

\large
\vspace*{0.5cm} 
{\noindent 
\section {\bf Comparison of Landau
Hydrodynamics with Experimental Rapidity Distributions}
}
\vspace{0.3cm}

We would like to compare the modified distribution with the Landau
distribution and experimental distributions for central AuAu
collisions at various energies \cite{Mur04,Ste05,Ste07}.  We can evaluate a
few quantities to get an idea of the differences.  Consider collisions
at $\sqrt{s_{_{NN}}}=200$ GeV.  The beam rapidity $y_b$ is $y_b=5.36$,
and the logarithm of the Lorentz contraction factor is $ L=4.67$.  The
difference between $y_b$ and $L$ is substantial and leads to different
shapes of the rapidity distributions as one observes in Fig. 6.

\begin{figure} [h]
\includegraphics[angle=0,scale=0.40]{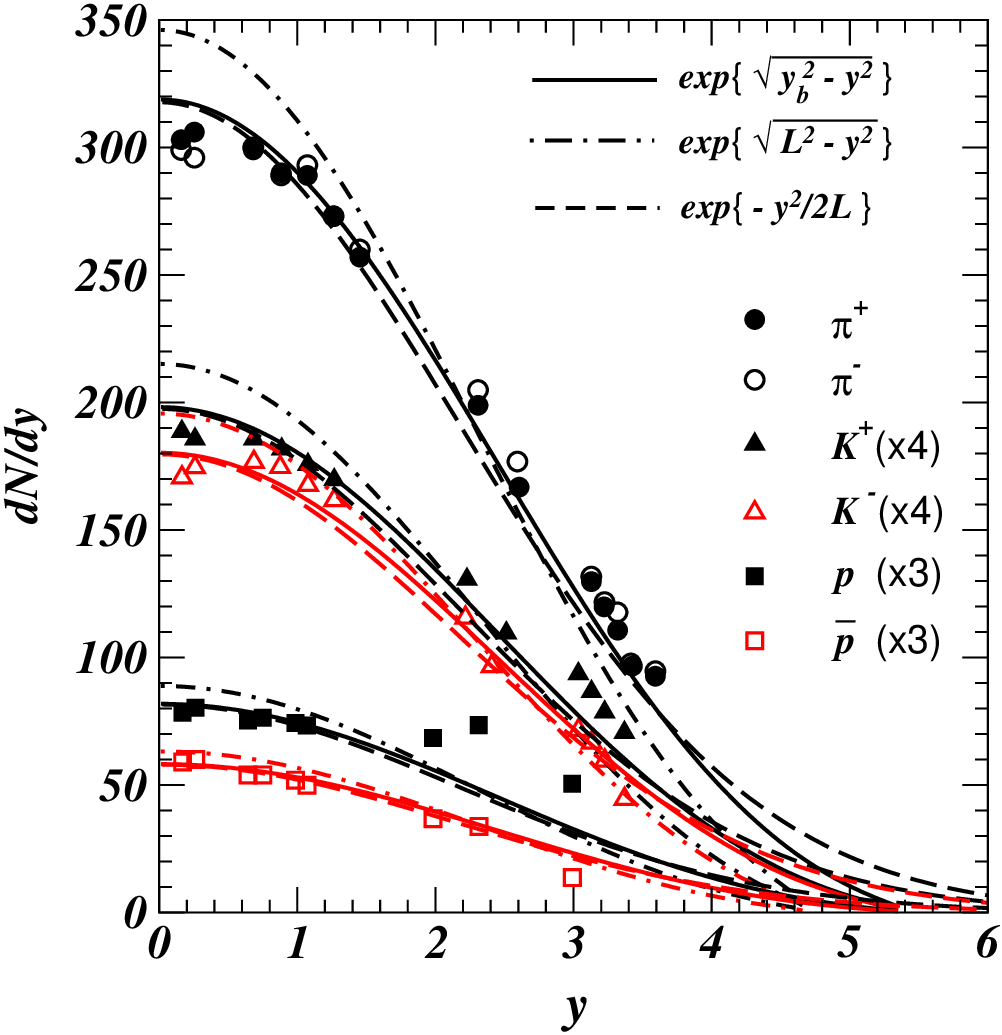}
\vspace*{0.0cm} 
\caption{ Comparison of experimental rapidity distribution with
  theoretical distribution in the form of $dN/dy \propto \exp\{
  \sqrt{y_b^2-y^2} \}$ (solid curves), Landau's distribution $dN/dy
  \propto \exp\{ \sqrt{L^2-y^2} \}$ (dashed-dot curves), and the
  Gaussian $dN/dy \propto \exp\{-y^2/2L \}$ (dashed curves) for
  produced particles with different masses. Data are from \cite{Mur04}
  for AuAu collisions at $\sqrt{s_{_{NN}}}=200$ GeV.  }
\end{figure}

Fig.\ 6 gives the theoretical and experimental rapidity distributions
for $\pi^+, \pi^-, K^+, K^-, p,$ and $\bar p$ \cite{Mur04}.  The solid
curves in Fig.\ 6 are the results for $\sqrt{s_{_{NN}}}=200$ GeV from
the modified distribution Eq.\ (\ref{new}) with the $y_b$ parameter,
whereas the dashed curves are the Landau distribution of Eq.\
(\ref{Lan}), $dN/dy \propto \exp\{\sqrt{L^2-y^2} \}$ with the $L$
parameter.  The theoretical distributions for different types of
particles have been obtained by keeping the functional forms of the
distribution and fitting a normalization constant to match the
experimental data.  We observe that Landau rapidity distributions are
significantly narrower than the experimental rapidity distributions,
whereas the modified distribution Eq.\ (\ref{new}) gives theoretical
results that agree better with experimental data.

As a further comparison, we show theoretical distributions calculated
with the Gaussian distribution of Eq.\ (\ref{gau}) as the dashed
curves.  We find that except for the region of large rapidities, the
Gaussian distributions is a good representation of the modified Landau
distribution.  The close similarity between the modified distribution
(\ref{new}) and the Gaussian distribution (\ref{gau}) explains the
puzzle mentioned in the Introduction.  The Gaussian distribution and
the original Landau distribution are different distributions.  Past
successes of the Gaussian distribution in explaining experimental
rapidity data \cite{Mur04,Ste05,Ste07} arises, not because it is an
approximation of the original Landau distribution (\ref{Lan}), but
because it is in fact close to the modified Landau distribution
(\ref{new}) that derives its support from a careful re-examination of
Landau hydrodynamics.

\begin{figure} [h]
\includegraphics[angle=0,scale=0.40]{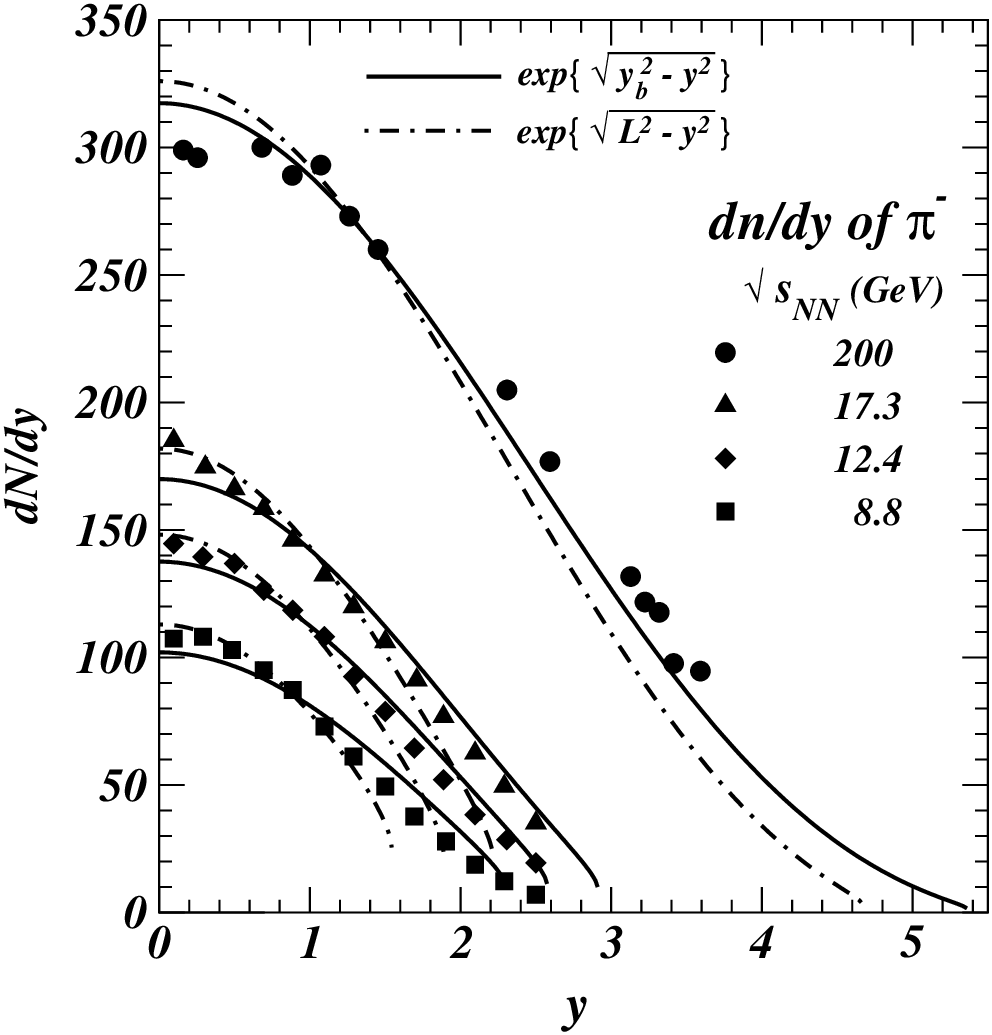}
\vspace*{0.0cm} 
\caption{ Comparison of experimental rapidity distribution with
  theoretical distribution in the form of $dN/dy \propto \exp\{
  \sqrt{y_b^2-y^2} \}$ (solid curves) and Landau's distribution $dN/dy
  \propto \exp\{ \sqrt{L^2-y^2} \}$ (dashed curves) for produced
  particles at different energies. Data are from the compilations in
  \cite{Mur04}.  }
\end{figure}

We compare theoretical distributions with the $\pi^-$ rapidity
distribution for collisions at various energies.  The solid curves in
Fig.\ 7 are the results from the modified distribution Eq.\ (\ref{new})
with the $y_b$ parameter, whereas the dashed curves are the Landau
distribution of Eq.\ (\ref{Lan}) with the $L$ parameter. The
experimental data are from the compilation of \cite{Mur04}.  The
modified distribution of Eq.\ (\ref{new}) appears to give a better
agreement with experimental data than the Landau distribution when all
the data points at all energies are considered.

While the comparison of the modified distribution gives a better
agreement with experimental data than the original Landau
distribution, it should be pointed out that the agreement can be
further improved.  Landau's condition of rapidity freezeout, Eq.\
(\ref{xtt}), is only a close estimate that already goes a long way in
providing a good description.  It may however admit refinement beyond
the first estimate.  We can in fact use the experimental data to guide
us for a better estimate of the rapidity freezeout condition which can
be conveniently modified to be
\begin{eqnarray}
 x(t_m) = \zeta a,
\end{eqnarray}
where $\zeta$ is of order unity but can allow small deviations from
unity.  In this case, the matching time becomes
\begin{eqnarray}
\label{refine1}
t_m(y) = 2 \sqrt{\zeta} a \cosh y,
\end{eqnarray}
and the rapidity distribution becomes
\begin{eqnarray}
\label{refine2}
dN/dy \propto \exp\{ \sqrt{[y_b+(\ln \zeta)/2]^2-y^2} \} .
\end{eqnarray}
The comparison in Figs.\ 6 and 7 for the $\pi^\pm$ distributions
indicates that a parameter $\zeta$ slightly greater than unity will
lead to a better agreement with experiment.

\vspace*{0.5cm}
{\noindent 

\section
{\bf Predictions of Particle Rapidity Distribution for LHC Energies}
\vspace{0.3cm}

We can re-write the rapidity distribution of charged
particles in terms of the normalized distribution $dF/dy$
\begin{eqnarray}
( dN_{ch} / dy) / (N_{\rm part}/2) = 
 [N_{ch} / (N_{\rm part}/2)] dF/dy. 
\end{eqnarray}
The normalized distribution $dF/dy$ is
\begin{eqnarray}
\label{shape}
\frac{dF}{dy} = \begin{cases}
A_{\rm norm} \exp\{\sqrt{y_b^2-y^2}\}   & {\rm for~ modified~ Landau~ distribution},\\
A_{\rm norm} \exp\{\sqrt{L^2-y^2}\}     & {\rm for~ Landau~ distribution},\\
A_{\rm norm} \exp\{{-y^2/2L}\}     & {\rm for~ Gaussian~Landau~distribution},
\end{cases}
\end{eqnarray}
where $A_{\rm norm}$ is a normalization constant such that
\begin{eqnarray}
\int  dF/dy =1.
\end{eqnarray}
As we remarked earlier, the approximate Landau hydrodynamical model is
restricted in the region of $|y|<y_b$ for the modified distribution
and $L<|y|$ for the original Landau distribution.  The neglect of the
tail region of the distribution will not affect significantly the
distribution and its normalization.

With the knowledge of the total charged multiplicity from Fig.\ 1, and
the shape of the rapidity distribution from Eq.\ (\ref{shape}), we can
calculate $dN_{\rm ch}/dy / (N_{\rm part}/2)$ as a function of
rapidity.  Fig. 5 gives the predicted rapidity distributions at LHC
energies.  For heavy-ion collisions at $\sqrt{ s_{_{NN}}}=$ 5.5 TeV
with full stopping ($\xi=1$), the maximum value of $dN/dy$ per
participant pair is about 22 at midrapidity.  For $pp$ collisions at
$\sqrt{ s_{_{NN}}}=$ 14 TeV with $\xi=0.5$, the maximum $dN/dy$ is
approximately 24 at $y=0$.  The widths of the rapidity distributions
are $\sigma_y\sim 3$.  The solid curves are for the modified
distribution, the dashed-dot curves are for the original Landau
distribution, and the dashed curves are for the Gaussian distribution.
These results provide useful theoretical data for comparison with
future experiments.

\begin{figure} [h]
\includegraphics[angle=0,scale=0.50]{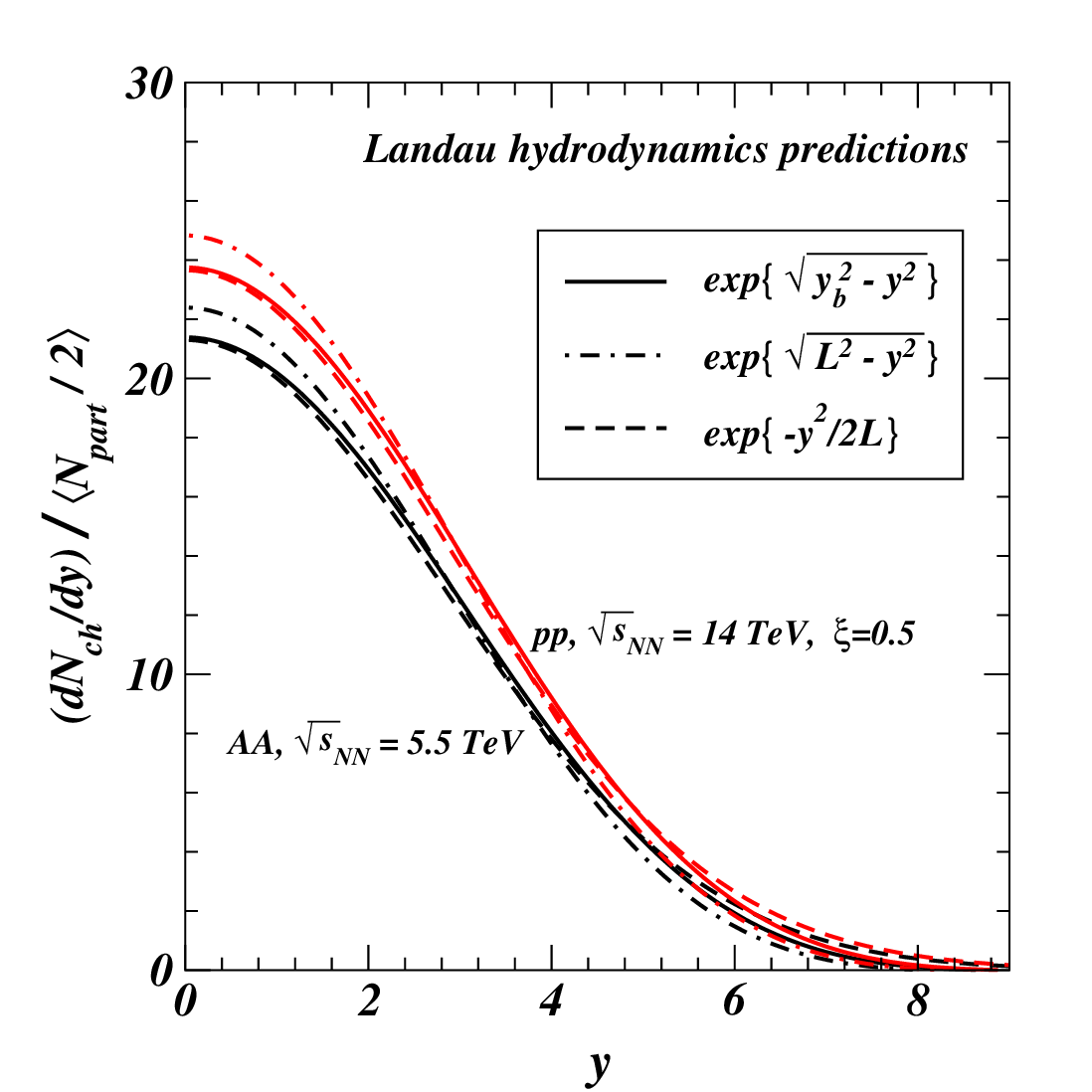}
\vspace*{0.0cm} 
\caption{The predicted rapidity distributions $dN_{\rm ch}/dy /
 (N_{\rm part}/2)$ of charged particles produced in (i) $pp$
 collisions at $\sqrt{s_{_{NN}}}=14$ TeV with $\xi=0.5$, and (ii) AA
 collisions at $\sqrt{s_{_{NN}}}= 5.5$ TeV with full stopping
 ($\xi=1$), in Landau hydrodynamics.  The solid curves are obtained
 with the modified distribution, the dashed-dot curves are obtained
 with the original Landau distribution, and the dashed curves with the
 Gaussian distribution.  }
\end{figure}

\section
{\bf Generalization to Non-Central Collisions}
\vspace{0.3cm}

We have considered so far the case of central collision, for which the
dimensions in all directions before Lorentz contraction are all equal.
In non-central collisions, the transverse radius $a_\phi/2$ will
depend on the azimuthal angle $\phi$ measured relative to the $x$ axis
as depicted in Fig. 1.

We can easily generalize the Landau model to the case of non-central
collisions.  The first stage of longitudinal expansion proceeds in the
same way.  The initial energy density $\epsilon_0$ will need to be
adjusted as it depends on the impact parameter.  The important
quantity is however the temporal dependence of the transverse
displacement.  Following the same Landau arguments as in the central
collision case, Eq.\ (\ref{displ}) for the transverse displacement can
be generalized to the non-central collision case to be
\begin{eqnarray}
\label{displrho}
\frac{4}{3} \epsilon (u^0)^2 \frac{ 2 \rho(\phi,t) }{ t^2}
= \frac{ \epsilon}{ 3a_\phi/2}. 
\end{eqnarray}
where $\rho(\phi,t)$ is the transverse displacement at azimuthal angle
$\phi$.  As a consequence, the transverse displacement depend on
$\phi$ and $t$ as
\begin{eqnarray}
 \rho(\phi, t) = \frac{t^2}{4 a_\phi (u^0)^2 }=\frac{t^2}{4 a_\phi
 \cosh^2 y} .
\end{eqnarray}
We can set up the Landau condition for the onset of the second stage
as the condition that the transverse displacement $\rho(\phi, t)$ is
equal to the transverse dimension $a_\phi$,
\begin{eqnarray}
 \rho(\phi, t_m) = a_\phi.
\end{eqnarray}
Thus, in the case of non-central collision, the Landau condition of
(\ref{xtt}) is changed to 
\begin{eqnarray}
t_m(y,\phi) = (a_\phi/a)\times 2 a \cosh y,
\end{eqnarray}
which shows that the rapidity freezeout time $t_m$ is reduced by the
factor $(a_\phi/a)$, compared to the freezeout time for central
collisions. For a fixed azimuthal angle $\phi$, fluid elements go from
the first stage motion to the second stage motion at different times.
The matching time $t_m$ (the rapidity freezeout time) occurs at
different instances for different rapidities.  Following the same
argument as before, Eq.\ (\ref{yyy}) for the non-central collision
case becomes
\begin{eqnarray}
y_{\pm} \bigl | _{t=t_m(y,\phi) }
= \ln (a_\phi/a)+ \ln(2a/\Delta_b)  \pm y,
\end{eqnarray}
where the longitudinal thickness of the
initial slab $\Delta_b$ depends on the impact parameter $b$.
As a consequence, the rapidity distribution for this non-central
collision is
\begin{eqnarray}
\frac{dN}{dy} &\propto& \exp\{\sqrt{\ln(2a/\Delta_b) + \ln (a_\phi/a)]^2-y^2} \}.
\end{eqnarray}

\large
\vspace*{0.5cm}
{\noindent 

\section {\bf 
Conclusions and Discussions}}
\vspace{0.3cm}

In many problems in high-energy collisions such as in the description
of the interaction of the jet or quarkonium with the produced dense
matter, it is desirable to have a realistic but simple description of
the evolution of the produced medium.  Landau hydrodynamics furnishes
such a tool for this purpose.

Recent successes of Landau hydrodynamics \cite{Mur04,Ste05,Ste07} in
explaining the rapidity distribution, total particle multiplicities,
and limiting fragmentation indicate that it contains promising degrees
of freedom.  Questions are however raised concerning the use of
pseudorapidity or rapidity variables, the approximate Gaussian form or
the square-root exponential form of the rapidity distribution, and the
values of the parameters in the rapidity distribution.

We start with the rapidity variable from the outset so that we do not
need to worry about the question of the rapidity or the pseudorapidity
variable.  We follow the formulation of the Landau hydrodynamics by
keeping careful track of the numerical constants that enter into the
derivation.  We confirm Landau's central results except that the
approximate rapidity distribution obtained by Landau needs to be
modified, when all numerical factors are carefully tracked.  In
particular, the rapidity distribution in the center-of-mass system
should be more appropriately given as $dN/dy \propto \exp \{
\sqrt{y_b^2-y^2}\}$, where $y_b$ is the beam nucleon rapidity, instead
of the Landau original result of $dN/dy({\rm Landau}) \propto \exp \{
\sqrt{L^2-y^2}\}$.  The modified distribution leads to a better
description of the experimental data, thereby supports the approximate
validity of Landau hydrodynamics as a description of the evolution of
the produced bulk matter.  Phenomenological fine-tuning can be further
introduced as in Eqs.\ (\ref{refine1}) and (\ref{refine2}) to relate
the rapidity freezeout time $t_m$ with the rapidity and transverse
radius.

The modified distribution differs only slightly from the Gaussian
distribution $dN/dy({\rm Gaussian}) \propto \exp \{ -y^2/2L\}$, that
has been used successfully and extensively in the literature
\cite{Mur04,Ste05,Ste07,Bus04,Car73}.  This explains the puzzle we
mention in the Introduction.  Even though the Gaussian Landau
distribution (\ref{gau}) is conceived as an approximate representation
of the original Landau distribution (\ref{Lan}) for the region of
small rapidity with $|y| \ll L$, it differs from the original Landau
distribution in other rapidity regions.  They are in fact different
distributions.  The Gaussian distribution has been successfully used
to explain experimental rapidity distribution data
\cite{Mur04,Ste05,Ste07}, not because it is an approximation of the
original Landau distribution (\ref{Lan}), but because it is in fact a
good representation of the modified Landau distribution (\ref{new})
that derives its support from a careful re-examination of Landau
hydrodynamics.  Thus, there is now a firmer theoretical support for
the Gaussian distribution (\ref{gau}) owing to its similarity to the
modified distribution of (\ref{new}).

The need to modify Landau's original distribution should not come as a
surprise, as the original distribution was intended to be qualitative.
Our desire to apply it quantitatively therefore lead to a more
stringent re-examination, with the result of the modification as we
suggest.

The quantitative successes of the modified Landau hydrodynamics make
it a useful tool for many problems in high-energy heavy-ion
collisions.  We can now view with some degree of confidence Landau's
picture for the evolution of the bulk matter.  The evolution first
proceeds with a one-dimensional longitudinal expansion with a
simultaneous transverse expansion.  Different parts of the fluid is
subject to different transverse displacements. Those in the central
rapidity receive the greatest transverse displacement and will freeze
out in $y$ at the earliest.  The final distribution is obtained by
collecting the bulk matter at different times of rapidity freezeout.

In spite of these successes, many problems will need to be examined to
make the Landau model an even better tool.  We can here outline a few
that will need our attention.  The distribution so far deals with flow
rapidity of the fluid elements, and the thermal distribution of the
particles inside the fluid element has not been included.  The folding
of the thermal distribution of the particles will broaden the rapidity
distribution and should be the subject of future
investigations. Another improvement is to work with a curvilinear
coordinate system in the transverse direction to obtain the transverse
displacement.  The will improve the description of the matching time
in the transverse direction.  One may wish to explore other forms of
the freezeout condition instead of Landau's transverse displacement
condition, to see how sensitive the results can depend on the
freezeout condition.  Finally, as the approximate  solution for the
one-dimensional is also available, it may also be of interest to see
how much improvement there can be in obtaining the matching time
estimates which enter into the rapidity freeze-out condition.

\vspace*{0.3cm} The author wishes to thank Prof. D. Blaschke for his
hospitality at the the Helmholtz International Summer School,
Bogoliubov Laboratory of Theoretical Physics, Dubna, Russia.  This
research was supported in part by the Division of Nuclear Physics,
U.S. Department of Energy, under Contract No.  DE-AC05-00OR22725,
managed by UT-Battelle, LLC.

\end{document}